\documentclass[
    ,final            
  ,numberedheadings 
  ]
  {aipproc}

\layoutstyle{6x9}
\newcommand{\etal}{\mbox{\it{et~al.}}}

\begin{document}

\title{Observations of Extrasolar Planets During the \\
non-Cryogenic Spitzer Space Telescope Mission}


\author{Drake Deming}{
  address={NASA's Goddard Space Flight Center, Planetary Systems Laboratory, \\
    Code 693, Greenbelt, MD 20771, USA}
}
\author{Eric Agol}{
  address={Department of Astronomy, University of Washington, Box 351580, Seattle, WA 98195-1580, USA}
}
\author{David Charbonneau}{
address={Department of Astronomy, Harvard University, 60 Garden St., MS-16, Cambridge, MA 02138, USA}
}
\author{Nicolas Cowan}{
  address={Department of Astronomy, University of Washington, Box 351580, Seattle, WA 98195-1580, USA}
}
\author{Heather Knutson}{
address={Department of Astronomy, Harvard University, 60 Garden St., MS-16, Cambridge, MA 02138, USA}
}
\author{Massimo Marengo}{
address={Harvard-Smithsonian Center for Astrophysics, 60 Garden St., MS-45, Cambridge, MA 02138, USA}
}
\begin{abstract}
Precision infrared photometry from Spitzer has enabled the first
direct studies of light from extrasolar planets, via observations at
secondary eclipse in transiting systems. Current Spitzer results
include the first longitudinal temperature map of an extrasolar
planet, and the first spectra of their atmospheres. Spitzer has also
measured a temperature and precise radius for the first transiting
Neptune-sized exoplanet, and is beginning to make precise transit
timing measurements to infer the existence of unseen low mass planets.
The lack of stellar limb darkening in the infrared facilitates precise
radius and transit timing measurements of transiting planets.  Warm
Spitzer will be capable of a precise radius measurement for
Earth-sized planets transiting nearby M-dwarfs, thereby constraining
their bulk composition.  It will continue to measure thermal emission
at secondary eclipse for transiting hot Jupiters, and be able to
distinguish between planets having broad band emission {\it vs.}
absorption spectra. It will also be able to measure the orbital phase
variation of thermal emission for close-in planets, even
non-transiting planets, and these measurements will be of special
interest for planets in eccentric orbits. Warm Spitzer will be a
significant complement to Kepler, particularly as regards transit
timing in the Kepler field.  In addition to studying close-in planets,
Warm Spitzer will have significant application in sensitive imaging
searches for young planets at relatively large angular separations
from their parent stars.

\end{abstract}

\maketitle

\section{Introduction}

The Spitzer Space Telescope (\citet{Werner2004}) was the first
facility to detect photons from known extrasolar planets
(\citet{Charbonneau2005, Deming2005}), inaugurating the current era
wherein planets orbiting other stars are being studied directly.
Cryogenic Spitzer has been a powerful facility for exoplanet
characterization, using all three of its instruments. Spitzer studies
have produced the first temperature map of an extrasolar planet
(\citet{Knutson2007a}), and the first spectra of their atmospheres
(\citet{Grillmair2007, Richardson2007}).  Spitzer will continue to
study exoplanets when its store of cryogen is exhausted.  `Warm
Spitzer' (commencing $\sim$ spring 2009) will remain at T $\sim$~35K
(passively cooled by radiation), allowing imaging photometry at 3.6
and 4.5 $\mu$m, at full sensitivity. The long observing times that are
projected for the warm mission will facilitate several pioneering
exoplanet studies not contemplated for the cryogenic mission.

\section{Extrasolar Planets in 2009}

Currently over 200 extrasolar planets are known, including 22
transiting planets (17 orbiting stars brighter than V=13). Some of
these have been discovered by the Doppler surveys, but an increasing
majority of the transiting systems are being discovered by
ground-based photometric surveys. However, the Doppler surveys remain
an efficient method to find hot Jupiters, and surveys such as N2K
(\citet{Fischer2005}) continue to be a productive source of both
transiting and non-transiting close-in exoplanets. The discovery rate
from the photometric surveys is accelerating, because these teams have
learned to efficiently identify and cull their transiting candidates,
and quickly eliminate false positives.  Several transit surveys (HAT,
TrES, and XO) recently announced multiple new giant transiting systems
(\citet{Burke2007, O'Donovan2007, Johns-Krull2007, Mandushev2007,
Bakos2007}), and a Neptune-sized planet has been discovered transiting
the M-dwarf GJ\,436 (\citet{Gillon2007a}). We estimate that the number
of bright (V<13) stars hosting transiting giant planets will increase
to $\sim$100 in the Warm Spitzer time frame.

The discovery of transits in GJ\,436b has stimulated interest in
finding more M-dwarf planets, both by Doppler surveys
(\citet{Butler2004}), and using new transit surveys targeted at bright
M-dwarfs.  It is reasonable to expect that $\sim$10 transiting hot
Neptunes will be discovered transiting bright M-dwarf stars by the
advent of the warm mission.  Moreover, the Doppler surveys are finding
planets orbiting evolved stars (\citet{Johnson2007}). The greater
luminosity of evolved stars can potentially super-heat their close-in
planets and facilitate follow-up by Warm Spitzer at 3.6 and 4.5~$\mu$m.

\section{Photometry Using Warm Spitzer}

Warm Spitzer has a particularly important role in follow up for bright
transiting exoplanet systems, as well as non-transiting systems,
because in 2009 it will be the largest aperture general-purpose
telescope in heliocentric orbit. Heliocentric orbit provides a
thermally stable environment, and it allows long periods of
observation, not blocked by the Earth. Although Kepler will have a
greater aperture than Spitzer, Kepler will be locked-in to a specific
field in Cygnus, so it cannot follow-up on the numerous bright
transiting systems that will be discovered across the sky.

The thermally stable environment of heliocentric orbit has proven to
be a boon for precision photometry from Spitzer. For example, the
recent Spitzer 8~$\mu$m observations of the HD\,189733b transit
reported by \citet{Knutson2007a}, illustrated in Figure~1, are among
the most precise transit observations ever made. These investigators
measured the planet-to-star radius ratio for HD\,189733b as 0.1545
$\pm$ 0.0002, corresponding to a precision of $\pm$~90 km in the
radius of the giant planet, and they also measured the orbital phase
variation of the planet's thermal emission.

\begin{figure}
\includegraphics[height=.4\textheight]{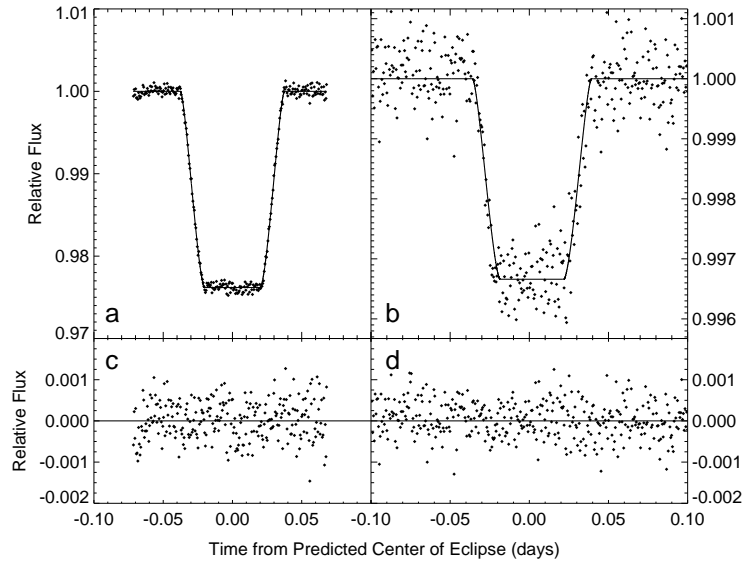}
\caption{Transit (left) and a $60\sigma$ secondary eclipse
detection (right) of HD\,189733b at 8~$\mu$m using a continuous
33-hour Spitzer photometry sequence (\citet{Knutson2007a}).}
\end{figure}

\section{Mass-Radius Relations}

Spitzer's precision for transits derives not only from its stable
thermal environment, but also from the lack of stellar limb darkening
in the IR. Without limb darkening, the transit becomes extremely
`box-like`, with a flat bottom (\citet{Richardson2006,
Knutson2007a}, see Figure~1). The IR transit depth yields the ratio of planet to
stellar area simply and directly, without the added uncertainty of
fitting to limb-darkening. Spitzer is now the facility of choice for
transiting planet radius measurements. A Warm Spitzer transit program
- exploiting the bright stellar flux at 3.6 and 4.5~$\mu$m - could
significantly improve our knowledge of the mass-radius relationship,
and clarify differences in bulk composition, for all but the faintest
hot Jupiter systems. Figure~2 shows the mass-radius relation for
several of the transiting giant planets (\citet{Charbonneau2006}). The
mass-radius relation encodes fundamental information on the global
structure of these planets. For example, HD\,149026b is inferred to
have a heavy element core of at least 70 Earth masses, based on the
small radius for its mass (Figure~2, and \citet{Sato2005}).  This
information is crucial to our understanding of planet formation, e.g.,
by the core accretion and gravitational instability mechanisms
(\citet{Lissauer2007}). The scientific utility of these measurements
will be maximized if all transiting exoplanet radii are measured to
high precision, in a mutually consistent manner.  Moreover, as the
Doppler and transit surveys discover Neptune to Earth-sized planets
orbiting M-dwarfs, the highest precision photometry will be needed to
measure their radii to a precision sufficient to constrain their
interior structure.

\begin{figure}
\includegraphics[height=.4\textheight]{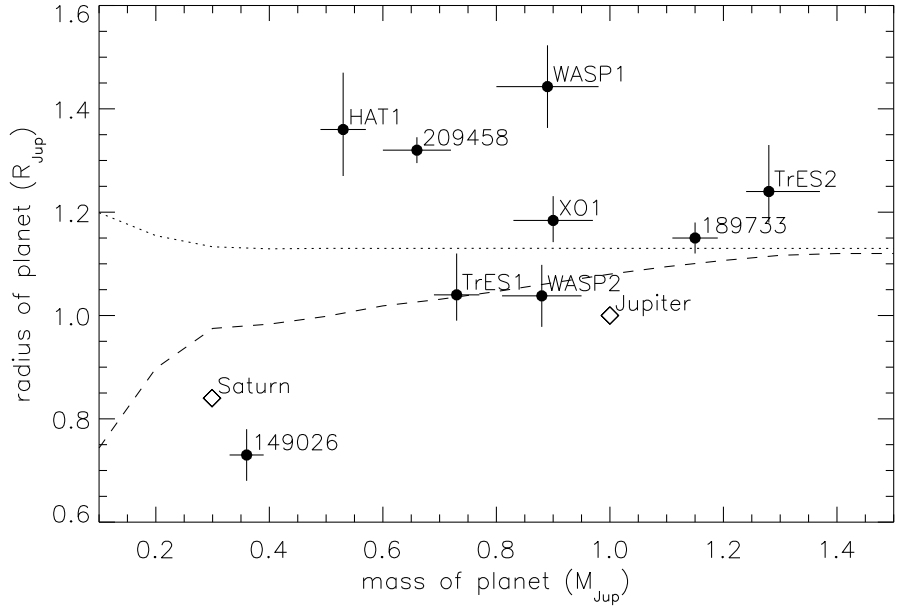}
\caption{Mass-radius relation for giant transiting exoplanets,
compared to Jupiter and Saturn (\citet{Charbonneau2006}). The lines
show the theoretical relations (\citet{Bodenheimer2003} for planets
having no core (dotted) and a 20 Earth-mass solid core (dashed).}
\end{figure}

\subsection{Spitzer vs. Ground-Based Photometry} 

Ground-based photometry in the z-band is achieving sub-milli-magnitude
levels of precision in many cases (\citet{Winn2007}), and can
determine the radii of some transiting giant planets to error limits
imposed by astrophysical uncertainty in the stellar mass. The most
favorable systems for ground-based observation are those occurring in
fields with numerous nearby reference stars of comparable brightness.
Planets transiting bright, spatially isolated, stars are not as
favorable for ground observation. Moreover, as the radius of the
transiting planet decreases, greater photometric precision is needed
to reach the limits imposed by uncertainty in the stellar mass.
Nearby M-dwarfs have flux peaks longward of the visible and z-band
spectral regions, and they often lack nearby comparison stars of
comparable infrared brightness.  Neptune- to Earth-sized planets
orbiting nearby M-dwarfs will therefore require infrared space-borne
photometry for the best possible radius precision. Figure~3
illustrates a single transit of a 1-Earth radius planet across an
M-dwarf, observed by Spitzer at 8~$\mu$m.  We simulated this case by
re-scaling a real case: Spitzer's recent photometry of GJ\,436b
(\citet{Deming2007a, Gillon2007b}).  Spitzer's nearly photon-limited
precision detects this Earth-sized planet to $7\sigma$ significance
in a single transit.

\begin{figure}
\includegraphics[height=.4\textheight]{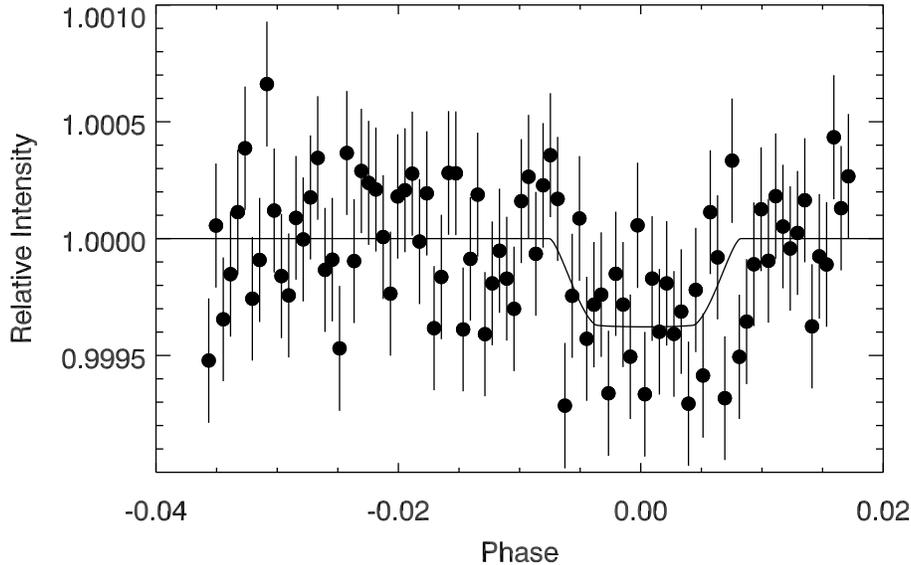}
\caption{A transit of a 1~Earth radius planet across an M-dwarf,
simulated by rescaling Spitzer 8~$\mu$m observations of GJ\,436b
(\citet{Deming2007a})}.
\end{figure}

Although Figure~3 is based on Spitzer observations at 8~$\mu$m, the
photon-limit for observations during the warm mission (e.g., at
4.5~$\mu$m) will be even more favorable, simply because stars are
brighter at the shorter wavelength. Stellar photometry at the
wavelengths used by the warm mission is affected by a pixel phase
effect in the IRAC instrument (\citet{Reach2005}), but that can be
successfully corrected by decorrelation (\citet{Charbonneau2005}), and
recent results have demonstrated secondary eclipse detections with a
precision of $\sim 10^{-4}$ (\citet{Charbonneau2005, Knutson2007b}).
The pixel phase effect should be correctable to even greater precison
using the large data sets contemplated for the warm mission.

\section{New Types of Transiting Planets}

Ongoing Doppler and transit monitoring of known hot Jupiters can
detect subtle deviations from Keplerian orbits
(\citet{Charbonneau2007}), indicating the presence of additional
planets, e.g., `warm Jupiters' in longer period orbits, or terrestrial
mass planets in low order mean motion resonances. The likely
co-alignment of orbital planes increases the chance those planets will
also transit, and intensive radial velocity monitoring could constrain
the transit time for giant planets. Warm Spitzer will be a sensitive
facility for confirming those transits, and extending the mass-radius
relation (Figure~2) to planets in more distant orbits, and even to
close-in terrestrial planets. Even lacking specific indications from
Doppler measurements, searches for close-in terrestrial planets in low
order mean motion resonances with known giant transiting planets
(\citet{Thommes2005}) are warranted using Warm Spitzer. These searches
could be combined with radius and transit timing measurements for the
giant planets, in the same observing program.

For stars not known to host a hot Jupiter, ongoing Doppler surveys and
space-borne transit surveys (e.g., COROT) will find more transiting
planets, extending to Neptune mass and below.  Spitzer transit
measurements can precisely determine the radii of small
planets. Exoplanet radius measurements and transit searches are
particularly appropriate for Warm Spitzer, because: a) stars are
bright in the 3.6 and 4.5 $\mu$m bands, while limb darkening is still
absent, b) stellar activity is muted at IR wavelengths, and c) longer
observing times are congruent with the goal of simplified operations
in the extended mission. Figure~4 shows a potential example of a
precise radius (allowing precise density) determination for a hot
super-Earth, compared to the mass-radius relation for solid exoplanets
of various composition (\citet{Seager2007}). In this case, the Spitzer
radius is sufficiently precise ($\pm$~0.1~Earth radii) to constrain
the bulk composition of this solid exoplanet by comparison to the
\citet{Seager2007} models.

\begin{figure}
\includegraphics[height=.4\textheight]{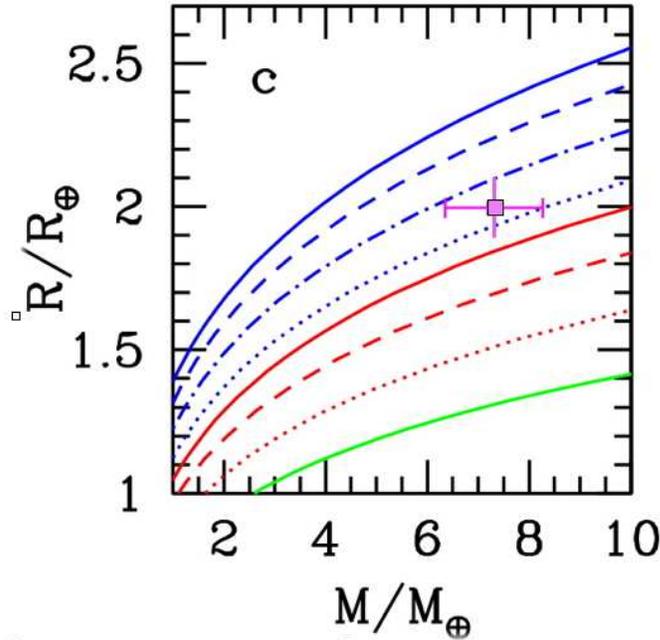}
\caption{Mass-radius relations for solid exoplanets of various
compositions, from \citet{Seager2007}. Blue represents water ice
planets, red are silicate planets, and green is a pure iron
planet. The magenta point is a hypothetical observation of a hot Earth
transiting an M-dwarf at 50 pc, observed by Warm Spitzer at 4.5
$\mu$m. The horizontal error bar is the mass error from Gliese 876d
(\citet{Rivera2005}); the vertical (radius) error bar is calculated
for Warm Spitzer, not including error in the stellar radius.}
\end{figure}

\section{Thermal Emission at 3.6 and 4.5 $\mu\rm{m}$}

\subsection{Absorption vs. Emission Spectra}

The secondary eclipse of a transiting planet has the greatest depth at
Spitzer's longest wavelengths. However, the eclipses are quite
detectable at Spitzer's shortest wavelengths, because these are close
to the peak of the Planck function at the temperatures of transiting
planets. Secondary eclipse photometry using Warm Spitzer can therefore
continue to define the brightness of exoplanets at 3.6 and 4.5 $\mu$m
for new bright transiting systems. This will be especially valuable if
complemented by ground-based detections exactly at the expected
spectral peaks at 2.2 and 3.8~$\mu$m, which is believed to be feasible
(\citet{Snellen2006, Deming2007b}).  The predicted spectrum of a hot
Jupiter exoplanet is illustrated in Figure~5, from
\citet{Charbonneau2005}.  Normally, the broad band spectrum is
expected to be shaped by water vapor absorption.  However, recent
Spitzer photometry of HD\,209458b (\citet{Knutson2007b}) indicates
much better agreement with a spectrum wherein the water bands appear
in {\it emission}, and can only be explained by the presence of a
thermal inversion at high altitude (\citet{Burrows2007}). The nature
of the high altitude absorber needed to create this inversion is
unknown, and may be connected to the specific properties of the
system, such as the planet's level of irradiation or surface
gravity. A comprehensive survey of the bright transiting systems would
make it possible to search for correlations between the presence of a
temperature inversion and other properties of the systems, thus
providing insight into the origin of the inversions. Such a survey
would also identify the best systems for spectroscopic follow up by
JWST.

The signature of such an inversion is easily observed in the 3.6 and
4.5~$\mu$m Spitzer bandpasses.  Standard models for atmospheres
without temperature inversions predict that the planets will appear
brighter at 3.6~$\mu$m than at 4.5~$\mu$m (see Figure~5) due to the
presence of water absorption bands at wavelengths longer than
4.5~$\mu$m.  With a thermal inversion the relative brightnesses in the
two bands are reversed, as the 3.6~$\mu$m flux is suppressed and the
4.5~$\mu$m flux is correspondingly enhanced by the presence of water
emission at the longer wavelengths.

\begin{figure}
\includegraphics[height=.4\textheight]{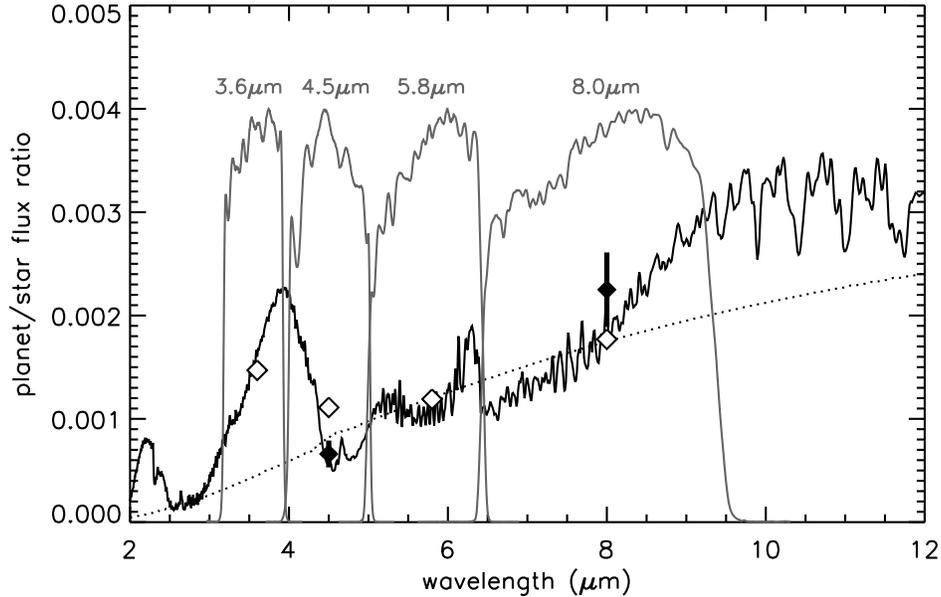}
\caption{Predicted specturm of a hot Jupiter exoplanet, shown in
comparison to observations of TrES-1 by \citet{Charbonneau2005} (solid
diamonds).  The expected contrast (planet divided by star) for the
IRAC bands is shown by open diamonds.  Note the higher contrast
expected at 3.6~$\mu$m compared to 4.5~$\mu$m, indicative of a water
absorption spectrum.  These contrast values will be reversed for a
water emission spectrum. The dotted line is a blackbody spectrum.}
\end{figure}

\subsection{Temporal Variations Due to Dynamics}

The cycle of variable stellar heating caused by planetary rotation can
drive a lively dynamics in hot Jupiter atmospheres
(\citet{Cooper2005}). In turn, the atmospheric dynamics will produce a
spatially and temporally varing temperature field, and the temperature
fluctuations are expected to be of large spatial scale for close-in
planets (\citet{Rauscher2007}). For the temperatures found in hot
Jupiter atmospheres (T$\sim$1200K), the Planck function at Warm
Spitzer wavelengths varies strongly with temperature. Consequently,
temporal variations in thermal emission from close-in planets should
be readily observable during the warm mission.  The secondary eclipse
depth of a given transiting system can vary, and observations of
multiple eclipses could yield key insights into the atmospheric
physics. Observations extended over a full orbit - even for
non-transiting systems - can potentially reveal variations in thermal
emission correlated with orbit phase (\citet{Cowan2007}). Orbital
phase variations can reflect the changing viewing geometry, but can
also be caused by strong atmospheric dynamics in response to the
variable stellar forcing that is characteristic of eccentric orbits.

Extrasolar planets have more eccentric orbits on average than do the
planets of our own solar system. In some cases, their eccentricity
extends to strikingly high values. For example, HD 80606b has an
eccentricity of 0.93 (\citet{Naef2001}).  During its close periastron
passage, it receives a stellar flux more than 1000 times greater than
the flux received by Earth from the Sun. This strong flux will cause a
rapid heating of the planet's atmosphere, and its time dependence
encodes crucial information on the radiative time scale, and thus the
composition, of the planet's atmosphere (\citet{Langton2007}).  This
rapid heating of HD 80606b and similar systems may be observable at
3.6 and 4.5 $\mu$m. In this regard it is interesting to note that the
exoplanet HD\,185269b orbits a sub-giant star, having greater than
solar luminosity, in a close orbit (6.8 day period), with an
eccentricity of 0.3 (\citet{Johnson2006}). The resultant strong
variation in stellar heating over the orbit will force a corresponding
variation in the planet's thermal emission, that should correlate with
orbital phase.  Recently, the transiting planets XO-3 and HAT-P-2 have
also been found to have a significant eccentricity
(\citet{Johns-Krull2007, Bakos2007}), opening the possibility to also
measure the spatial distribution of time dependent heating on the
planet's disk (\citet{Williams2006, Knutson2007a}). Given the high
precision possible from Spitzer, it may be possible to observe all of
these effects at 4.5~$\mu$m using Warm Spitzer.

\section{Transit Timing}

There is considerable recent interest in the indirect detection of
extrasolar terrestrial planets via their perturbations to the transit
times of giant transiting planets (\citet{Agol2005, Holman2005,
Steffen2005, Agol2007}). Spitzer transit photometry during the warm
mission is an excellent way to make precise transit timing
measurements. The lack of IR limb darkening is again a significant
advantage, because it results in very steep ingress and egress curves,
producing excellent timing precision.  \citet{Knutson2007a} found a
timing precision for the HD\,189733b transit of 6~seconds. Hubble
transit timing errors range from 10 to 50 seconds (\citet{Agol2007}),
although a 3~second timing precision was recently achieved for
HD\,189733b by \citet{Pont2007}. However, Spitzer has the advantage
that the lower contrast of star spots and plage in the IR as compared
to visible wavelengths minimizes systematic errors due to stellar
activity noise. Also, Spitzer's heliocentric orbit permits continuous
measurements before, during, and after transit - unlike Hubble where
blocking by Earth interrupts transits. Moreover, Spitzer transit timing
precision should be even better at the shorter wavelengths available
for the warm mission, because stars are brighter at shorter
wavelengths, and limb darkening remains negligible.

\begin{figure}
\includegraphics[height=.4\textheight]{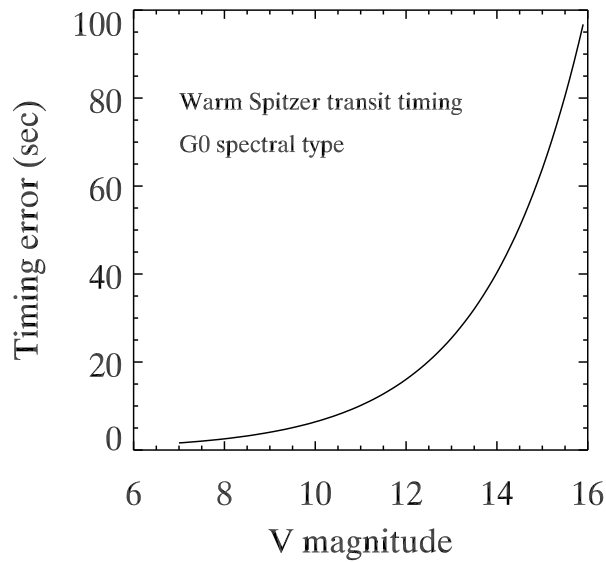}
\caption{Calculation of the transit timing error (1$\sigma$) for
Warm Spitzer observing giant planet transits at 4.5~$\mu$m, as a
function of stellar brightness.}
\end{figure}

The continued success of the ground-based transit surveys, the advent
of COROT, and the upcoming launch of Kepler, will provide a wealth of
targets for Warm Spitzer follow-up. We have calculated the transit timing
precision by Warm Spitzer at 3.6 and 4.5 $\mu$m, for solar-type stars
at different distances (Figure~6). This calculation is consistent with
the \citet{Knutson2007a} result, and it projects a Spitzer timing
precision of better than 40 seconds down to the faint end of Kepler's
range at V=14. This is sufficient to detect perturbations by
terrestrial planets well below one Earth mass in resonant orbits (Agol
et al. 2005), or to $\sim$ 10 (150 days/$P_1$) Earth masses for $P_2/P_1 <
4$, where $P_{1,2}$ are the periods of the transiting and perturbing planets
(\citet{Holman2005, Agol2005}). The Warm Spitzer mission will begin
at about the same time that Kepler begins to discover multiple new giant
transiting planets (February 2009 launch). The observing cadence of the
Kepler mission is not optimized for transit timing
(\citet{Basri2005}), so transit timing observations in the Kepler
field by Warm Spitzer could leverage and enhance Kepler's science
return.

\section{Direct Imaging}

Although the bulk of Spitzer's results for exoplanets have relied on
time series photometry and spectroscopy, Spitzer's high sensitivity in
imaging mode is also important for exoplanet imaging studies. Radial velocity
surveys have reached the precision required to detect planets only in
the last 10 years, so little is known about the frequency of
exoplanets and other low mass companions at distances greater than
$\sim$ 5 - 10 AU. This is unfortunate because determining the presence
of planetary mass bodies in the periphery of known exoplanetary
systems has important implications for their evolution. These include
studying the dynamical ``heating'' of the orbits in the system, which
may result in higher eccentricities (or even expulsion) of some
components, or in enhanced collision rates between the bodies in
extrasolar Kuiper belts, which may be responsible for the formation of
transient debris disks.

Imaging can in principle fill this observational gap. Three objects of
near planetary mass have already been detected within 300 AU from the
primary around the brown dwarf 2M1207 (\citet{Chauvin2004}) and the
stars GQ Lup (\citet{Neuhauser2005}) and AB Pic (\citet{Chauvin2005})
with ground based adaptive optics observations. Other optical and
near-IR searches from the ground and from space, have so far produced
negative results.

\begin{figure}
\includegraphics[height=.4\textheight,angle=-90]{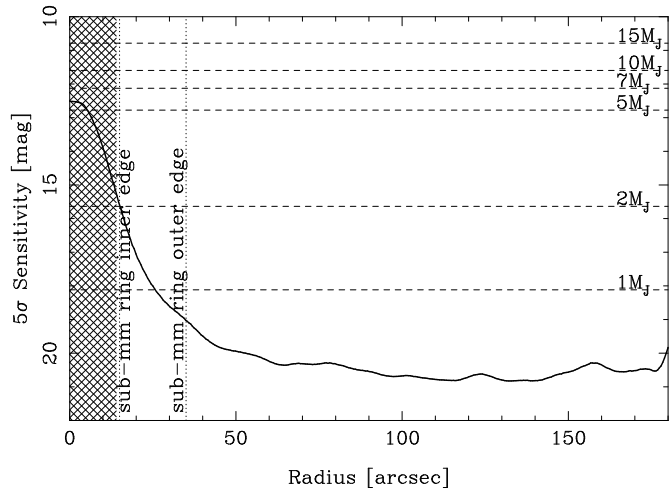}
\caption{Residual noise, after PSF subtraction, of 4.5 $\mu$m $\epsilon$
~Eridani radial profile (\citet{Marengo2006}). Dashed lines are the
predicted fluxes (from \citet{Burrows2003}) of 1 Gyr old planets with
mass from 1 to 15 M$_J$. Dotted lines enclose the region where the
sub-millimeter debris disk around this star is located.}
\end{figure}

Warm Spitzer may provide a significant contribution in this arena, as
the two surviving IRAC bands are particularly suited for the detection
of cool extrasolar planets and brown dwarf companions (T and the
so-called Y dwarfs). A gap in molecular opacities of giant planets
near 4.5~$\mu$m allows emission from deep, warmer atmospheric layers
to escape: giant planets are very bright at this wavelength
(\citet{Burrows2003}). A strong methane absorption band strongly
depresses the planetary flux at 3.6~$\mu$m: as a result, the IRAC
[3.6]-[4.5] color of planetary mass bodies is expected to be unique,
and allow for their identification among background objects in the
field. Model atmosphere calculations predict that a 1~Gyr old, 2~M$_J$
planet around a star 10 pc from the Sun will have a 4.5~$\mu$m
magnitude of $\sim$~18. Such planets are detectable today with IRAC
provided that the diffracted light from the central bright star can be
removed at the planet's image location. The stability of Spitzer's
optical and pointing systems assures that the stellar Point Spread
Function (PSF) is highly reproducible, allowing much fainter nearby
sources to be identified in the PSF wings using differential
measurements.

This search has already been carried out for the debris disk star
$\epsilon$~Eridani, which is also home of a Jovian class radial
velocity planet orbiting the star at 3.4~AU (\citet{Hatzes2000}). The
search has set stringent limits for the mass of external planetary
bodies in the system (including the area occupied by the debris disk,
\citet{Marengo2006}), and demonstrated that this technique is
sensitive to the detection of planets with mass as low as 1~M$_J$
(Figure~7) outside the 14 arcsec radius (50~AU) where the IRAC frames
are saturated. The search radius can be reduced to $\sim 5$ arcsec or
less by using shorter frame times available in IRAC subarray mode. A
pilot search of 16 nearby stellar systems is being conducted in
Spitzer cycles 3 and 4. These programs will identify possible candidates
based on their [3.6]-[4.5] colors, which will need to be verified by
second epoch observations during the warm mission, to detect their
common proper motion with the primary.

The Spitzer warm mission will provide the opportunity to extend the
search of planetary mass companions through imaging techniques to a
large number of systems in the solar neighborhood, probing a search
radius from $\sim$ 10 to 10,000 AU around stars within 30 pc from the
Sun. This search will be sensitive to masses as low as a few Jupiter
masses, depending on the age and distance of the systems. These
observations will be complementary to ground based radial velocity and
imaging searches with adaptive optics systems, given the larger field
of view and higher sensitivity of IRAC, in a wavelength range where
the required contrast ratio (as low as $10^{-5}$ of the parent star flux)
is more accessible than in the optical and near-IR.

\begin{theacknowledgments}

We thank the Spitzer Science Center for the opportunity to consider
and discuss the potential for exoplanet science during the warm
mission.  We are grateful to Josh Winn and Andy Gould for helpful
conversations and remarks regarding the relative merits of of
ground-based vs. space-borne photometry.  We also acknowledge
informative conversations with Greg Laughlin on the effects of heating
in eccentric orbits.

\end{theacknowledgments}

\bibliographystyle{aipproc}   

\end{document}